\documentclass[A4]{dem8}

\usepackage[utf8]{inputenc}

\usepackage{graphicx}
\usepackage{subfig}
\usepackage{amsmath}
\usepackage{amsfonts}
\usepackage{amssymb}
\usepackage{color}
\usepackage{listings}
\usepackage{textcomp}

\definecolor{codegreen}{rgb}{0,0.6,0}
\definecolor{codegray}{rgb}{0.5,0.5,0.5}
\definecolor{codepurple}{rgb}{0.58,0,0.82}
\definecolor{backcolour}{rgb}{0.95,0.95,0.92}

\lstdefinestyle{mystyle}{
    backgroundcolor=\color{backcolour},   
    commentstyle=\color{codegreen},
    keywordstyle=\color{magenta},
    numberstyle=\tiny\color{codegray},
    stringstyle=\color{codepurple},
    basicstyle=\footnotesize,
    breakatwhitespace=false,         
    breaklines=true,                 
    captionpos=b,                    
    keepspaces=true,                 
    numbers=left,                    
    numbersep=5pt,                  
    showspaces=false,                
    showstringspaces=false,
    showtabs=false,                  
    tabsize=2
}
 
\newcommand{\MESA}{MESA-PD}

\title{A Modular and Extensible Software Architecture for Particle Dynamics}

\author{Sebastian Eibl$^1$, Ulrich Rüde$^{1, 2}$}

\heading{A Modular and Extensible Software Architecture for Particle Dynamics}

\address{$^{1}$ Lehrstuhl f\"ur Informatik 10 (Systemsimulation)\\
Friedrich-Alexander-Universit\"at Erlangen-N\"urnberg, Cauerstr. 11, 91052 Erlangen, Germany
\and
$^2$ CERFACS\\
42 Avenue Gaspard Coriolis, 31057 Toulouse, Cedex 01, France
}

\keywords{Particle Dynamics, High Performance Computing, Code Generation, Parallel Computing}

\abstract{
Creating a highly parallel and flexible discrete element software requires an interdisciplinary approach, where expertise from different disciplines is combined. On the one hand domain specialists provide interaction models between particles. On the other hand high-performance computing specialists optimize the code to achieve good performance on different hardware architectures. In particular, the software must be carefully crafted to achieve good scaling on massively parallel supercomputers. Combining all this in a flexible and extensible, widely usable software is a challenging task.

In this article we outline the design decisions and concepts of a newly developed particle dynamics code \MESA\ that is implemented as part of the waLBerla multi-physics framework. Extensibility, flexibility, but also performance and scalability are primary design goals for the new software framework. In particular, the new modular architecture is designed such that physical models can be modified and extended by domain scientists without understanding all details of the parallel computing functionality and the underlying distributed data structures that are needed to achieve good performance on current supercomputer architectures. This goal is achieved by combining the high performance simulation framework waLBerla with code generation techniques. All code and the code generator are released as open source under GPLv3 within the publicly available waLBerla framework (www.walberla.net).
}

\begin{document}

\section{Motivation}
\label{sec:motivation}
Modern simulation software frameworks must fulfill a variety of requirements. In a systematic design process, these requirements must be identified and formulated clearly in the preparatory stages of the development process. Only then a software design can be created that will fulfill as many requirements as possible. Here we will follow this approach in the design of a new simulation software for particle dynamics. Such a simulation software should be easy to use and it should be extensible by the end user. At the same time, high performance is desirable and a parallelization suitable for current supercomputers is necessary. Also, maintainability and portability are essential features since the software framework will often be used for a long time, exceeding the life span of computer systems. In this paper, we will discuss some of the requirements that we have identified for our software framework. We explain the implications and our design decisions to satisfy the requirements. In particular, we try to identify and modularize different components of particle dynamics simulations such that specialists can work on them independently. In the following, we present three technical requirements that guide the development of "Modular and Extensible Software Architecture for Particle Dynamics" (\MESA).

\begin{itemize}
\item \textbf{flexible domain partitioning} Current state-of-the-art parallelization of particle dynamics software uses spatial domain partitioning. This domain partitioning is crucial for an efficient parallel execution. The way the domain is partitioned has impact on how well the workload can be balanced between the different processes and how much communication is needed during synchronization. Since different simulation scenarios will lead to different optimal domain partitionings, a flexible approach is needed to achieve good performance in all cases. A flexible domain partitioning is also important for coupled simulations which run in parallel. Ideally all coupled simulation components must share the same partitioning to avoid communication overheads. Therefore it is a great advantage when the domain partitioning can be easily adapted to that of other simulation modules.

\item \textbf{extensible particle data structure} Different interaction models between the particles require the particle to have certain properties. These can be material parameters, electric charge, temperature, and many more. Also, the framework user might want to store additional information when extending the functionality. However, all these possible properties will not be needed for every simulation. The general conventional implementation would have to provide all possible properties in the particle data structure. In most applications, however, only few features are used, therefore, many properties are unused. This leads to a potentially huge amount of memory that is wasted and that slows down the simulation if it has to be copied or sent over the network. Also, the maintainability of the particle data structure gets harder due to its size and dependencies. Here, a flexible approach is needed that allows to add and remove individual particle properties as needed for every specific simulation scenario.

\item \textbf{interaction model} \MESA\ is intended to be used as a scientific research tool. This makes it essential that particle interaction models can be adapted. It should be possible to implement new models easily within a short time to be able to test different approaches (\emph{rapid prototyping}). In the best case, domain specialists who are not profoundly familiar with the framework should be able to modify existing or develop new interaction models. To achieve this, the interaction models must be decoupled from the rest of the framework as much as possible. Note, however, that this can easily lead to a design conflict regarding optimization. The possibility should still exist for a computer scientist to optimize the source code of the interaction model to achieve the best possible performance.
\end{itemize}

After identifying these requirements we will discuss our approach to fulfill them in Sec.~\ref{sec:domain_partitioning}--\ref{sec:kernel}.

\section{Related Work}
Different approaches were taken to create flexible, modular, performant and highly parallel software frameworks for particle simulations. 
The \emph{LIGGGHTS}\footnote{https://www.cfdem.com/} DEM software package uses input files in its own custom language to describe the simulation. It also offers the possibility to extend the software framework by writing own extensions. However, these extensions must be written in C++ and some understanding of the underlying structure of the software framework is needed. 
Similarly \emph{GranOO}\footnote{https://www.yakuru.fr/granoo/index.html} uses XML based input files to set up and run the simulation. The simulation is comprised of so called "PlugIns" which take care of different parts of the simulation. These PlugIns can be selected by the user via the XML file. It is also possible for the user to add custom PlugIns written in C++ or Python. These PlugIns access the simulation via a special API.
\emph{YADE}\footnote{https://yade-dev.gitlab.io/trunk/} goes one step further and uses the scripting language Python for its input files. This allows the user the greatest flexibility in setting up the simulation as the full potential of Python and its packages can be used.

Another approach to let the user extend the software framework is shown by some molecular dynamics packages. They use high level languages like Python or their own embedded domain specific language (DSL) to describe the simulation. In order to achieve good performance on various hardware architectures they use just-in-time compilation to generate user specific executables for various architectures. However, to support MPI or CUDA additional wrapper libraries like pyMPI and PyCUDA are needed. Many packages using this technique claim that this can be done with almost no loss in performance compared to native C++ code. Packages that provide such capabilities with a varying degree of just-in-time compilation are for example \emph{OpenMM}~\cite{Eastman2017}, \emph{HOOMD-blue}~\cite{Anderson2008}, \emph{MDL}~\cite{Cickovski2007} and \emph{ppmd}~\cite{Saunders2018}.

\section{Our Contribution}
The plugin model to extend an existing framework is limited by the API provided by the framework. This indirect access to the simulation data may be a limiting factor if data is needed that cannot be accessed via the API. Since the API cannot be extended by the plugin itself either workarounds must be implemented or the extension is not possible. Depending on the internal data structures of the simulation framework, the interaction may be slow if data has to be converted between the framework and the plugin.

\begin{figure}[h]
  \centering
  \includegraphics[width=1.0\textwidth]{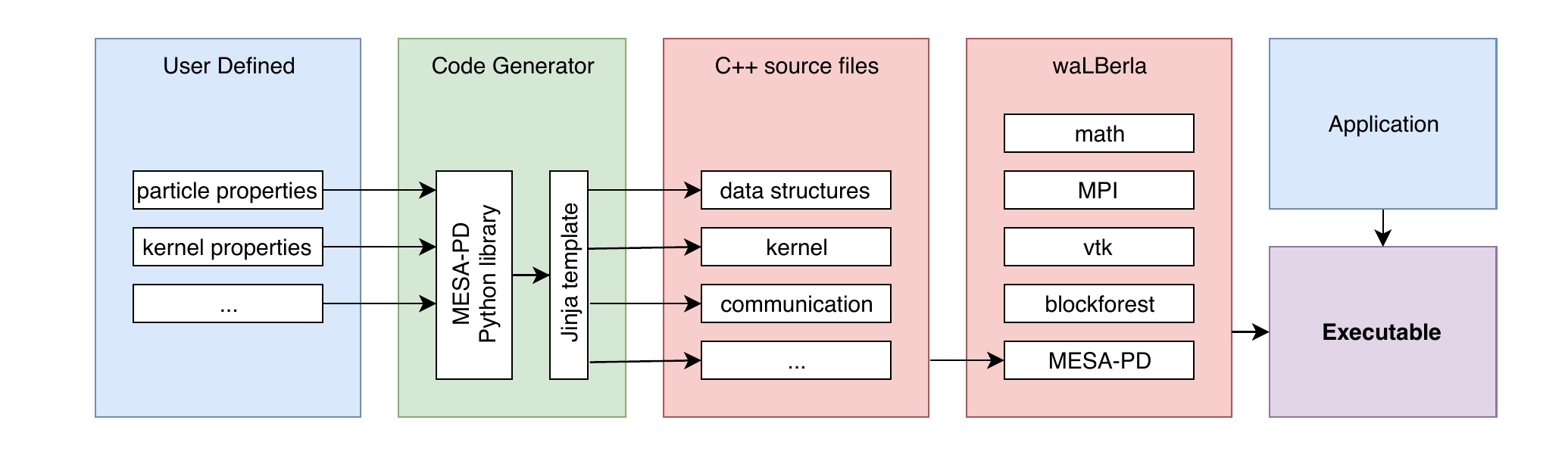}
  \caption{Workflow of the code generation in the new \MESA\ module. The code generation starts with the user specifying what is needed for the simulation. This information is specified with a high level Python library. The library then forwards the information to the correct Jinja templates for further processing. The Jinja template engine then creates the C++ source files of the \MESA\ waLBerla module. After this generation process, \MESA\ will be compiled and the application can be linked with the waLBerla modules to produce the final executable.}
  \label{fig:workflow}
\end{figure}

In the \MESA\ design we therefore favor a single C++ codebase with no indirections. We use the waLBerla multi-physics framework~\cite{Godenschwager2013} as a basis for the implementation of \MESA. waLBerla is an open source high-performance framework written in C++ that has shown excellent performance and scalability for Lattice Boltzmann~\cite{Schornbaum2016,Schornbaum2017} and particle dynamics simulations~\cite{Preclik2015,Eibl2018}. Using this framework as a starting point, gives instant access to many utility functions like storing results into SQLite databases, writing vtk output files, etc. Additionally, advanced functionality like load balancing~\cite{Eibl2018a} can be used. We extend this framework with the new \MESA\ module which is the result of the requirements formulated in Sec.~\ref{sec:motivation}. Writing a native C++ module requires deep knowledge about the framework as well as profound programming skills. To achieve our requirements, however, we want the user to be able to make changes quickly. There are some proposals for the future C++ standard like reflections, metaclasses, and concepts which can make writing modules easier. Unfortunately they are not available in the current C++ standard. We, therefore, introduce an additional code generation step which generates parts of the module automatically. This additional step of compile-time transformations extends the possibilities the programmer has over bare C++ code. The code generation itself is performed via Jinja templates\footnote{http://jinja.pocoo.org/}. Jinja templates allow us to introduce placeholders in the source code which will be filled later with the correct piece of code. But Jinja templates not only allow simple placeholder but also control structures like \emph{if} and \emph{for}. With these control structures one can disable replacements as well as generating a new line of source code for every item in a list. This alone, however, is not enough to make the coding much simpler. On top of the Jinja templates we introduce a Python library. The purpose of this Python library is to collect information about the simulation the user wants to conduct in a single place and then forward the information to all templates that work on that piece of information. In the end this allows the user of the framework to specify the simulation properties in a very high level representation. The distribution of the information is then handled by the Python library and the low level C++ source code is automatically generated by the Jinja template engine. This workflow is visualized in Fig.~\ref{fig:workflow}.

The code generation has to be run once by the user before the application gets compiled. With this approach we aim to leverage the full potential of a highly optimized framework and a unified code basis. Simultaneously, we can profit from the increased flexibility and straightforwardness gained from the code generation approach.

In the rest of this paper we will discuss the implementation satisfying the requirements presented. We will also point out where the new code generation approach greatly simplifies the programming task. A flexible domain partitioning approach using an interface is presented in Sec.~\ref{sec:domain_partitioning}. The handling of particle data and the extension with user supplied data is discussed in Sec.~\ref{sec:data_structure}. Finally we present our kernel interface in Sec.~\ref{sec:kernel}. This allows to introduce new kernels without any deeper knowledge about the rest of the framework.

\section{Flexible Domain Partitioning}
\label{sec:domain_partitioning}
\begin{figure}[h]
  \centering
  \subfloat[domain partitioning into regular blocks used by the forest of octrees]{\includegraphics[width = 0.3\textwidth]{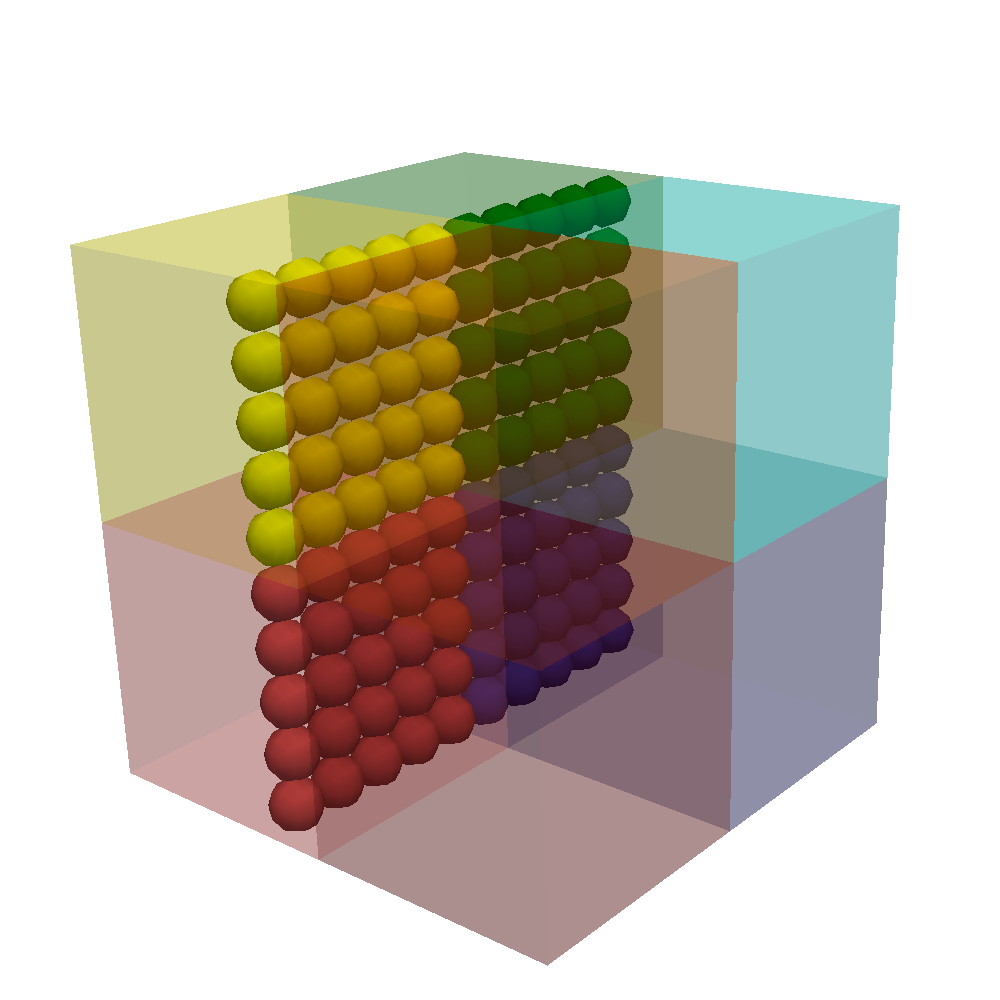}}
  \hfill
  \subfloat[tetrahedral domain partitioning used by the HyTeG framework]{\includegraphics[width = 0.3\textwidth]{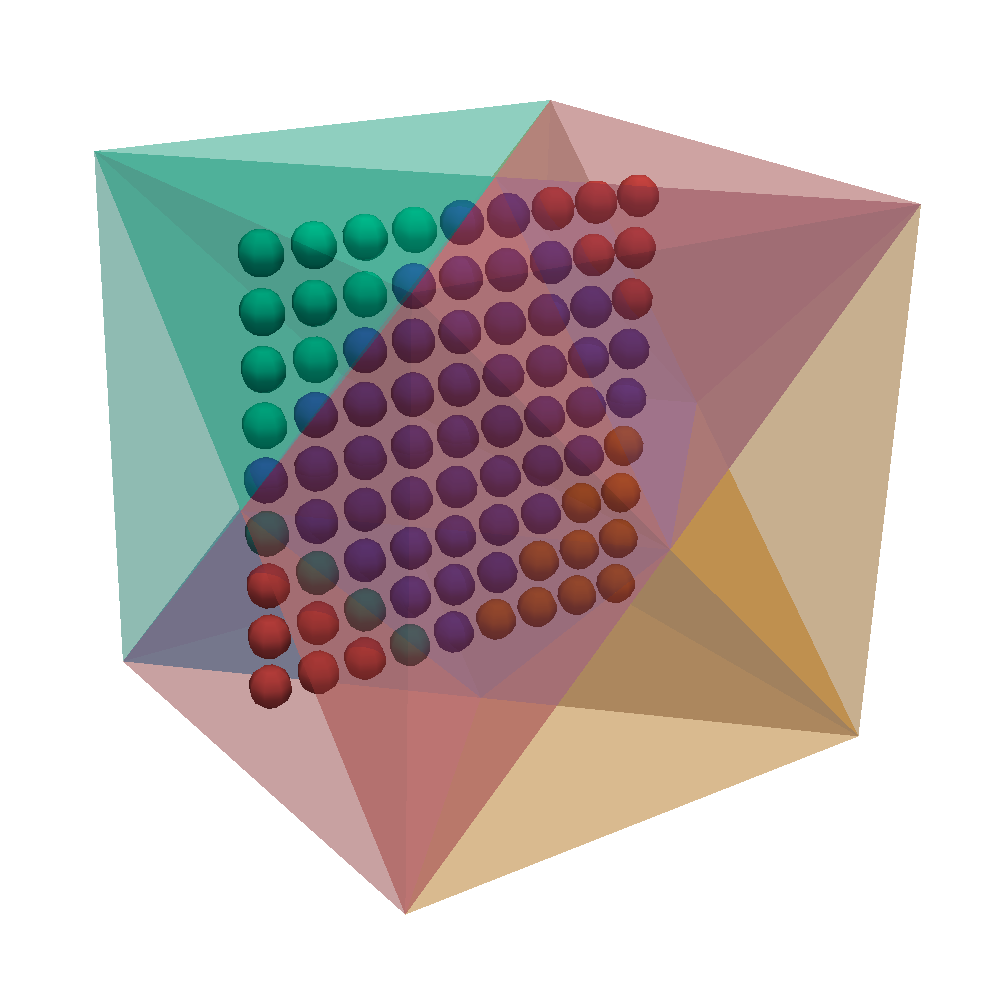}}
  \hfill
  \subfloat[artificial domain partitioning into spherical shells]{\includegraphics[width = 0.3\textwidth]{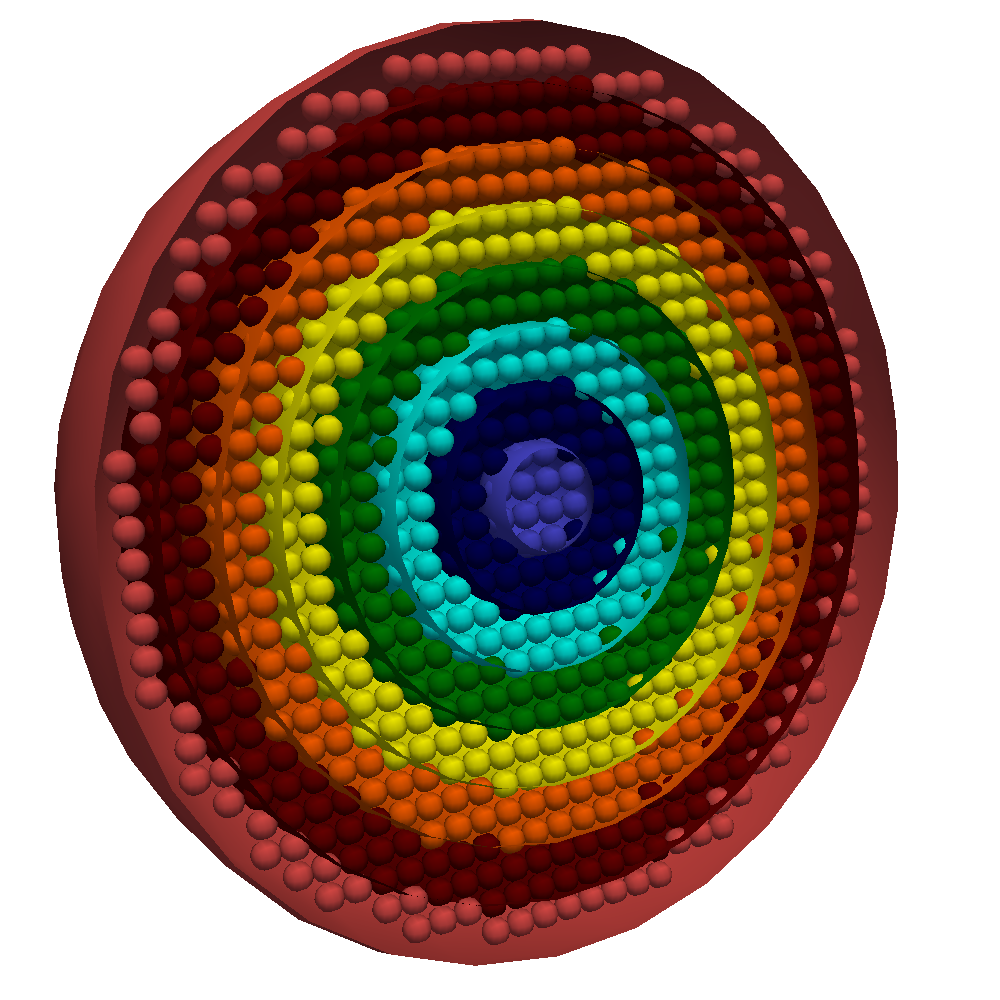}}
  \caption{This series of illustrations shows various domain partitionings for particle simulations. The particles are colored according to the MPI rank they belong to. The different subdomains are also colored accordingly. All domain partitionings can be used in a parallel simulation without changing the particle dynamics code. Only the domain interface has to be implemented for every partitioning.}
  \label{fig:DomainPartitioning}
\end{figure}

When running simulations in parallel with a spatial domain partitioning approach each process is responsible for a specific subpart of the simulation domain. All particle interactions within this region can be easily detected and resolved since the process has all necessary information. However, at subdomain interfaces, the available information is insufficient. Particles from neighboring subdomains might overlap with the local subdomain but information about the particle is not locally present. Typically this problem is solved by introducing \emph{ghost particles}. Ghost particles are copies of particles located at other processes which do not belong to the subdomain of the process. In particle dynamics simulations where interactions occur as soon as particles collide, ghost particles are created when they overlap the subdomain. These ghost particles are created and updated by the synchronization algorithm. State-of-the-art synchronization algorithms, as described in \cite{Rapaport1991,Eibl2018}, are independent of the actual subdomain geometry. The information they need is the ownership of a particle, i.e., which process stores the data, which is typically the process which handles the subdomain the geometrical center of the particle lies in. And additionally, if the particle overlaps adjacent subdomains which is commonly done by checking the bounding volume \cite{Schinner_1999,Perkins_2001} against the neighboring subdomains. So all tasks can be reduced to geometric functions that check points and bounding volumes against subdomains. As long as one can provide these functions the synchronization algorithm can be implemented without specific knowledge about the domain partitioning. For this purpose \MESA\ introduces an interface to the domain partitioning that covers all functionality that is needed for an efficient parallelization of the simulation. Thus the algorithms become independent of the domain partitioning. Different implementations of this interface can take care of various peculiarities of the simulation carried out. Three exemplary domain partitionings realized with this interface can be seen in Fig.~\ref{fig:DomainPartitioning}. Within the waLBerla multi-physics framework an implementation for the distributed forest of octrees domain partitioning~\cite{Schornbaum2016,Schornbaum2017} is available. This allows an easy interaction with additional waLBerla modules like the Lattice Boltzmann Method implementation within a single application~\cite{Rettinger2017a,Rettinger2019}. The interface is also implemented by the HyTeG finite elements multigrid framework which uses unstructured tetrahedral meshes for the domain partitioning~\cite{Kohl_2018}. An artificial domain partitioning into spherical shells is also shown in Fig.~\ref{fig:DomainPartitioning}.

\section{Particle Data Structure}
\label{sec:data_structure}
A flexible particle data structure that can be adapted according to the needs of the current simulation scenario is important to keep the memory footprint small and performance high. However, changing the properties of the particles also involves adapting some algorithms. For example, the algorithm which updates all ghost particles needs to know what particle properties it should synchronize. To simplify the process of adding and removing particle properties and to update all algorithms accordingly we use our additional code generation library. The library offers high level functions to add properties to the particle data structure. This greatly reduces the workload for the programmer and it is also less demanding on the programming skills. A new particle property can be added in the following form:
\begin{lstlisting}[language=Python]
addProperty(name, datatype, defValue, syncMode)
\end{lstlisting}
\texttt{name} names the property and \texttt{datatype} specifies the data type of the property. The data type can be any valid C++ data type. The \texttt{defValue} is the value the property gets initialized with when a new particle is created. The most interesting parameter for parallel simulations is the \texttt{syncMode}. This parameter controls when and how this property gets synchronized between different processes. Different modes are available, namely:

\textbf{NEVER}
This property gets never synchronized. This is useful if it is used only to store intermediate results.

\textbf{COPY}
This property is copied exactly once when a new ghost particle is created. This is typically used when the property does not change but is different for every particle. Depending on the simulation this can be something like \emph{mass}, \emph{particle radius}, etc.

\textbf{MIGRATION}
Properties annotated with this syncMode are synchronized when the ownership of a particle changes (i.e. the particle leaves the current subdomain and is now in the subdomain of a different process). This can be used for example to synchronize the old force in a velocity verlet integration scheme. 

\textbf{ALWAYS}
This property is synchronized in every iteration. This is used for properties which change frequently like \emph{position}.

During the code generation, the library passes this information to the templates that need it. The templates are then translated into C++ source code. In the following we want to illustrate this process. First, the user defines that the particle data structure should contain three properties: position, radius and force. The position should be synchronized in every synchronization step whereas the radius only needs to be synchronized when a new (ghost) particle is created. The force is recalculated in every time step and is therefore never synchronized. All this information is specified with the following three lines:
\begin{lstlisting}[language=Python]
addProperty("Position", "Vec3", "0,0,0", "ALWAYS")
addProperty("Radius", "double", "0", "COPY")
addProperty("Force", "Vec3", "0,0,0", "NEVER")
\end{lstlisting}
This information is then forwarded to all source code templates. An exemplary template that packs the particle \texttt{particle} into a buffer looks like this:
\begin{lstlisting}[language=Python]
{%- for prop in properties %}
{%- if prop.syncMode in ["COPY", "ALWAYS"] %}
buf << particle.get{{prop.name}}();
{%- endif %}
{%- endfor %}
\end{lstlisting}
The for loop prints the enclosed source code for every property into the C++ file. Additionally, the if statement selects only specific properties that are needed in this context. With the information provided by the user the template gets expanded into:
\begin{lstlisting}[language=C++]
buf << particle.getPosition();
buf << particle.getRadius();
\end{lstlisting}

One of the major benefits of this approach is that the user only has to specify the particle properties once. According to what the user specified, multiple source files as well as many occurrences within one source file are adapted. This greatly reduces the burden of remembering all places in the source code which need to be adapted to work with this particular set of properties. It is also less error prone since pairs like packing and unpacking are both generated. This eliminates possible inconsistencies. But not only all algorithms are adapted automatically also the user gets exactly the particle data structure which perfectly suits the scenario. However, one additional step is needed. Before compiling the application code the user has to run the code generation once. After that the application can be compiled like usual. 

\section{Particle Interaction Models}
\label{sec:kernel}
In this section, we discuss how particle interactions can be implemented in \MESA. Saunders et al.~\cite{Saunders2018} used the fact that most operations in their molecular dynamics code are carried out either on every molecule (apply gravity, do time integration) or on every pair of molecules (interactions). They proposed to separate these operations into individual functions which they called \emph{kernels}. With this approach one can isolate the interaction models from the rest of the code. The idea that all operations concerning molecules can be written as kernels can also be applied to general particle dynamics codes. Using this concept, a domain specialist can implement an interaction model by writing a function which takes two particles as input and calculates the interaction. In our approach we go even further and also decouple the kernel code from the actual data. We use a so called \emph{accessor} interface that maps between the kernels and the actual data structure. When a kernel accesses particle properties it does so via the accessor. The accessor than locates the data and passes it to the kernel. The accessor is used for reading as well as writing particle data. This allows to switch the implementation of the accessor interface without touching the kernels. With this approach one can change the particle data structure independently of the kernels. Only the implementation of the accessor interface has to be adapted. 

\begin{lstlisting}[language=C++, caption=Example of an Euler integration kernel. All accesses to particle properties are handled by the Accessor template. This way the kernel is completely independent of the data structure and can be used with whatever data structure as long as an appropriate accessor implementation is available.]
template <typename Accessor>
inline void EulerIntegrator::operator()(const size_t idx,
                                        Accessor& ac) const
{
   ac.setPosition(idx, ac.getInvMass(idx) * ac.getForce(idx) * dt_ * dt_ + 
                       ac.getLinearVelocity(idx) * dt_ + 
                       ac.getPosition(idx));
   ac.setLinearVelocity(idx, ac.getInvMass(idx) * ac.getForce(idx) * dt_ + 
                             ac.getLinearVelocity(idx));
   ac.setForce         (idx, Vec3(0, 0, 0));
}
\end{lstlisting}

This approach comes with many benefits. First of all, domain specialists who write the kernel code can do so without worrying about where data is stored and how to access it. All kernel accesses to the outside world are represented as \emph{give me this information} and \emph{store that information}. This way the kernel code is completely independent of the rest of the framework. It is also possible to use the kernels with a different framework as long as the data structures of the framework can be accessed via a particle accessor interface making the kernels widely usable. This greatly increases the flexibility of the kernels and also offers more possibilities in coupling different simulation frameworks.

\section{Conclusion}
In this paper we presented a new approach to extend the high-performance framework waLBerla with a new particle dynamics module \MESA. \MESA\ employs a code generation step to simplify the task of writing modules for highly optimized frameworks. This additional step is realized by using a combination of Jinja templates and a newly designed Python library. With this approach, the user can give a high level description of the simulation using the Python library. In a second code generation step C++ source files are created by the Jinja template engine using the information the user has provided. This way the C++ source code files are tailored exactly to the description the user has given.

For the design of the module we have identified requirements that are essential for a modern particle dynamics framework. We then presented our resolution of these requirements within our newly developed module. The new module allows a more flexible domain partitioning in parallel simulations. This simplifies the task of coupling simulations as well as experimenting with more efficient domain partitionings tailored for a specific situation. We also introduced an advanced approach to create individual data structures for every simulation without manual code rewrites. Finally, we showed our approach to decouple the code for particle interactions not only from the rest of the framework but also make it independent of the data structures used to store the properties of the particles. 

The new design has many benefits for the user of the framework. The code generation approach greatly reduces the lines of code the user has to write himself. If a single piece of information is needed at multiple places throughout the source code the code generation takes care of adapting all files accordingly. For example, after defining a particle property with just one line in Python the correct packing and unpacking functions to MPI buffers, debug output to the terminal, vtk output, output to databases, etc. are generated automatically. It not only saves time for the user it also ensures that there are no inconsistencies between the functions which could possibly lead to hard to track down errors.

The strict separation of data and kernels via the accessor interface allows to change parts of the code without interfering with other parts. With this approach specialists do not have to know the whole code base to introduce their knowledge. Additionally, due to the clear separation of all source code parts they might also be transferable to other frameworks.

\bibliographystyle{ieeetr}
\small
\bibliography{literature}

\end{document}